\documentstyle[eqsecnum,preprint,prb,aps]{revtex}
\begin{document}
\title {
Quantum Hall Effect \\
in  Three-dimensional Field-Induced Spin Density Wave Phases\\
with a Tilted Magnetic Field}
%
\draft
\author {Yasumasa Hasegawa}
\address {Faculty of Science, \\
Himeji Institute of Technology, \\
Kamigouri-chou, Akou-gun, Hyogo 678-12, Japan}

\date {\today}
\maketitle


\begin{abstract}
The quantum Hall effect  in the three-dimensional anisotropic
tight-binding electrons is investigated in
the field-induced spin density wave phases
with a magnetic field
tilted to any direction. The Hall conductivity,
 $\sigma_{xy}$ and $\sigma_{xz}$, are shown to be quantized
as a function of the wave vector of FISDW,
while $\sigma_{yz}$ stays
zero, where $x$ is the most conducting direction
and $y$ and $z$ are
perpendicular to $x$.

\end{abstract}

\section{Introduction}
Quantum Hall effect has been observed in
organic conductors,
(TMTSF)$_2$PF$_6$ and (TMTSF)$_2$ClO$_4$
\cite{Naughton88,Cooper89,Hannahs89}
in the field-induced spin density
wave (FISDW) phases.
  These
organic conductors are quasi-one-dimensional conductors and
described
by a tight binding Hamiltonian
with anisotropic hopping matrix elements,
$t_b / t_a \approx 0.1$ and $t_c / t_a \approx 0.003$.
The Fermi
surface consists of two warped planes in
the absence of a magnetic field.
If the
nesting of the Fermi surface is perfect,
the staggered susceptibility
corresponding to the nesting vector
which connects two planes of the
Fermi surface diverges as temperature becomes low.
In such case the spin density wave (SDW) is stabilized by the
repulsive interaction of electrons.
The nesting of the Fermi surface
is imperfect in general due to transverse hoppings. For the
imperfect nesting case the transition temperature of SDW depends on
the imperfectness and the strength of the electron interaction.
Since the transverse hoppings are enhanced by pressure,  the SDW
state is suppressed  as the pressure is increased.
The divergence of the staggered susceptibility recovers when the
magnetic field is applied \cite{GorkovLebed84}, and the successive
transitions to FISDWs occur as
the magnetic field is
increased
\cite{Montambaux85,Yamaji85,Virosztek86,Yamaji87,Osada92}.
One can explain the stability of FISDW by the fact that the
energy is lowered when a gap at the Fermi surface is opened due
to the combined effect of the electron interactions and magnetic
field.

  A magnetic
field required to be one flux quantum, $\phi_0=hc_0/e$, per
electron in each plane for these organic conductors is extremely
large ($\sim 10^4$ Tesla).
In other words the quantum Hall effect in organic
conductors is observed in  much smaller field than
that is necessary to
fill electrons only in the lowest few Landau levels.
Moreover, the
Hall conductivity is observed to change its sign
in some region of the
magnetic field \cite{Ribault84}.  The Quantum Hall effect in the
presence of FISDW has been
studied theoretically
\cite{Poilblanc87,Kohmoto90,Yakovenko91,Hasegawa93pre}.
The quantization of
the Hall conductivity is explained by using the  St\v{r}eda formula
\cite{Poilblanc87} or  the general theory of the Hall effect in the
periodic system \cite{Kohmoto90,Yakovenko91,Hasegawa93pre}.
These authors studied the case
that the magnetic field is perpendicular
to the conducting plane.

Due to the hoppings along the third direction new phenomena are
expected  when the magnetic field is tilted \cite{Lebed86}. In
three-dimensional lattice there exist three fluxes,
$\phi_a$, $\phi_b$ and $\phi_c$ per unit area in each plane, while
only one flux per unit cell is present in two-dimension.
 Indeed, many cusps of the
magnetoresistance have  been observed as a function of
a direction of the magnetic field \cite{Osada91,Naughton91}.
FISDW in three-dimension in the tilted magnetic field or
in non-orthogonal lattice has been studied
\cite{MontambauxL89,ChenMaki89,SunMaki93}. The Hall
conductivity $\sigma_{xy}$ and $\sigma_{xz}$ are predicted to be
quantized. These values are argued to be fractionally
quantized \cite{MontambauxL89,MontamK90}.
These authors, however,  studied  the case that the magnetic
field is perpendicular to one of the axis, say $a$, when two fluxes
$\phi_b$, and $\phi_c$ exist instead of three
in the general direction of the field.

On the other hand the general formula for the quantum Hall effect
in the periodic system in three-dimension was given by
Halperin\cite{Halperin87} and Kohmoto {\it et al.}
\cite{KohmotoHW92}. They have shown that the  conductivity
tensor is quantized when the Fermi energy is in the energy gap.

In this paper we study the Hall conductivity in the
anisotropic three-dimensional lattice  with the
magnetic field tilted to any direction.
In section~\ref{noninteracting}
the conductivity tensor is calculated for the non-interacting
electrons in the orthorhombic lattice. The quantized value
is given in the perturbation in $t_b / t_a$ and $t_c / t_a$. The
quantum Hall effect in the presence of  FISDW is given in
section~\ref{fisdw}. The generalization to the triclinic lattice is
done in section~\ref{triclinic}.

\section{Non-interacting case}
\label{noninteracting}
In this section we study the
 anisotropic tight-binding electrons on the orthorhombic
lattice in the magnetic field,
\begin{equation}
 {\cal H}_0 = -t_a \sum_{(i,j)_a,\sigma} e^{i \phi_{ij}}
c_{i,\sigma}^{\dagger}
           c_{j,\sigma}
       - t_b \sum_{(i,j)_b,\sigma} e^{i \phi_{ij}}
c_{i,\sigma}^{\dagger}
         c_{j,\sigma}
        - t_c \sum_{(i,j)_c,\sigma} e^{i \phi_{ij}}
c_{i,\sigma}^{\dagger}
         c_{j,\sigma}
\end{equation}
where $c_{i,\sigma}^{\dagger}$ and $c_{i,\sigma}$ are
creation and annihilation operators of electron with spin
$\sigma$ at site $i$, $t_a$, $t_b$
and $t_c$ $( t_a \gg t_b$, $t_c )$
are the hopping matrix elements along $a$,  $b$ and $c$ directions,
respectively, and \begin{equation}
\phi_{ij} = {2\pi \over \phi_0} \int_i^j {\bf A} \cdot d{\bf l},
\end{equation}
where $\bf A$ is a vector potential.
 In this section  and section~\ref{fisdw}
the $a$, $b$ and $c$ axes
are assumed to be orthogonal each other and
parallel to the $x$, $y$ and $z$
direction, respectively.

The uniform magnetic field, ${\bf B}$, is applied in any direction,
\begin{equation}
  {\bf B} = (B_x, B_y, B_z).
\end{equation}
The fluxes through the unit area
perpendicular to the $a$, $b$ and $c$ axes are given
by
\begin{mathletters}
\begin{equation}
  \phi_{x} =  b c B_x ,
\end{equation}
\begin{equation}
  \phi_{y} =  c a B_y ,
\end{equation}
\begin{equation}
  \phi_{z} =  a b B_z ,
\end{equation}
\end{mathletters}
where $a$, $b$, and $c$ are the lattice constants.

First we consider a rational flux case, {\it i.e.},
the flux through a unit area in each plane is a rational number,
$(\phi_x/\phi_0, \phi_y/\phi_0, \phi_z/\phi_0) =
( p_x/q_x, p_y/q_y, p_z/q_z)$  with mutually prime integers
$p_{\alpha}$ and $q_{\alpha}$
($\alpha = x,y$ or $z$). Then
the Hamiltonian in the presence of the magnetic field is described
as a generalization of the problem in two dimension studied by
Azbel\cite{Azbel} and Hofstadter\cite{Hofstadter} to three
dimension
\cite{MontamK90,Hasegawa90,KunsztZee91,Hasegawa92,HKM93}.
The energy spectrum for the rational flux case can be obtained
numerically as eigenvalue of Harper equations. The  size
of the Harper equations is  the least common multiple of
$q_x$, $q_y$ and $q_z$, which we define $Q$.
The volume of the magnetic Brillouin zone is $1/Q$ of the Brillouin
zone in the absence of a magnetic field. For  a fixed momentum
${\bf k}$ in the magnetic Brillouin zone there exist $Q$ eigenvalues
and $Q$ eigenstates. By varying the momentum we get $Q$ bands.
In two dimension these bands do not overlap.
Although the bands may overlap in
three dimension,
  it has been shown that  the energy
gaps due to a magnetic field exist in the wide range of
parameters\cite{Hasegawa90,Hasegawa92}.
We can use the extended zone by unfolding the magnetic Brillouin
zone to the Brillouin zone in the absence of a field. Then the
energy is obtained uniquely for each momentum.

 The quantum Hall effect is expected
in the three-dimensional case when some of the energy bands are
completely filled and the other bands are
empty
\cite{Halperin87,MontambauxL89,MontamK90,KohmotoHW92}.
The  conductivity
tensor for a filled band is given as
\cite{Thouless82,Halperin87,KohmotoHW92}
 \begin{equation}
\sigma_{\alpha \beta}
 =
2 {e^2 \over h } {1 \over 4 \pi^2 i}
 \int  d^3k
 \left[
  {\partial \over \partial k_{\alpha}}
 \left(
 < \Psi |
  {\partial   \over \partial k_{\beta}}  |\Psi >
 \right)
     -
  {\partial   \over \partial k_{\beta}}
 \left(
 < \Psi |
  {\partial \over \partial k_\alpha}  | \Psi >
 \right)
 \right]  .
\label{HallConductivity}
\end{equation}
where the factor of 2 comes from the spin degrees of freedom,
$\alpha$ and $\beta$ are $x$, $y$, or $z$, $| \Psi >$ is the wave
function for the filled band and the integral is performed in  the
magnetic Brillouin zone.
Kohmoto, Halperin and Wu\cite{KohmotoHW92}
have shown that  if the Fermi energy lies in the energy gap, the
conductivity tensor is described as
 \begin{equation}
\sigma_{\alpha \beta} = 2 {e^2 \over 2 \pi h} \varepsilon_{\alpha
\beta \gamma} G_{\gamma},
\label{sigma1}
\end{equation}
with the vector in the reciprocal lattice,
\begin{equation}
{\bf G} =  - (l_{a'} {\bf G}_{a'} + l_{b'} {\bf G}_{b'} + l_{c'}
{\bf G}_{c'})
\end{equation}
where
${\bf G}_{a'}$, ${\bf G}_{b'}$ and ${\bf G}_{c'}$ are the
fundamental reciprocal vectors satisfying ${\bf B}
\cdot {\bf G}_{a'} = {\bf B} \cdot {\bf G}_{b'} = 0$
and $l_{a'}$, $l_{b'}$ and $l_{c'}$ are integers.
These integers are the first Chern numbers on the tori obtained by
slicing the three-torus of the magnetic Brillouin zone.
Since we can take any fundamental reciprocal vectors, ${\bf G}$ is
written as
\begin{equation}
{\bf G} = - (l_a {\bf G}_a + l_b {\bf G}_b + l_c {\bf G}_c),
\label{defG2}
\end{equation}
where  ${\bf G}_a = (2\pi / a) {\bf x} /|{\bf x}|$,
${\bf G}_b = (2\pi / b) {\bf y} / |{\bf y}|$, ${\bf G}_c
= (2\pi / c) {\bf z} / |{\bf z}|$ for the orthorhombic lattice and
$l_{a}$, $l_{b}$ and $l_{c}$ are integers.  If the direction of the
magnetic field is changed infinitesimally, {\bf G} stays constant
as long as the Fermi energy lies in the energy gap.
 Although ${\bf G}_{a'}$, ${\bf G}_{b'}$ and ${\bf G}_{c'}$
depend on the direction of the magnetic field,
${\bf G}_{a}$, ${\bf G}_{b}$ and ${\bf G}_{c}$ do not depend on the
magnetic field. As a result, while $l_{a'}$,  $l_{b'}$ and $l_{c'}$
are not constant,  $l_{a}$,  $l_{b}$ and $l_{c}$ are constant for
the infinitesimal change of the magnetic field.

If the Fermi energy lies between the $r$th and $r+1$th band from
the bottom of the energy, the conductivity tensor is given as the
summation of the contributions from filled $n$ bands.
The contributions from the filled bands cancel each other except
that from the upper energy gap of the $r$th band\cite{Kohmoto89}.

In order to calculate the Hall conductivity explicitly,
  we take the vector potential ${\bf A}$ as
\begin{equation}
 {\bf A} = (0,\  (B_z x - B_x z),\  -
{B_y \over B_z} (B_z x - B_x z)).
\label{VectorPotential}
 \end{equation}
In the above we have assumed $B_z \neq 0$ without loss of
generality, since we may exchange $y$ and $z$ in the case of $B_z =
0$.   With this vector potential the non-interacting Hamiltonian is
written \begin{eqnarray}
   {\cal H}_0 =& &- t_a \sum_{\bf k, \sigma} 2 \cos (a k_x)
c^{\dagger}_{\sigma}  ({\bf k}) c_{\sigma} ({\bf k})
 \nonumber \\
  & & -t_b
  \sum_{\bf k,\sigma} e^{-i b k_y}  c^{\dagger}_{\sigma}
 ({\bf k} - {\bf u} ) c_{\sigma}({\bf k})  + h.c.
 \nonumber \\
  & & -t_c e^{ i \pi {\phi_x \phi_y \over \phi_0 \phi_z}}
\sum_{\bf k,\sigma}  e^{-i c k_z}  c^{\dagger}_{\sigma}
 ({\bf k} +  {\bf u}') c_{\sigma}({\bf k})  + h.c. ,
\end{eqnarray}
where
$h.c.$ means Hermitian conjugate,
\begin{equation}
 {\bf u} = \left( {2\pi \over a} {\phi_z \over \phi_0} , 0,
-{2\pi \over c} {\phi_x \over \phi_0}
\right) ,
\label{VectorU}
\end{equation}
and
\begin{equation}
 {\bf u}' =  {\phi_y \over \phi_z} {\bf u} .
\label{VectorUp}
\end{equation}
We define the state as
\begin{equation}
  \psi_{\sigma} ({\bf k})  = c^{\dagger}_{\sigma} ({\bf k}) | 0 > ,
\end{equation}
with vacuum $| 0 >$.
The Hamiltonian mixes $ \psi_{\sigma} ({\bf k})$
with $ \psi_{\sigma} ({\bf k} \pm {\bf u}) $ and
$  \psi_{\sigma} ({\bf k} \pm  {\bf u}') $.
If $\phi_x = 0$, $k_y$ and $k_z$ are constants of motion.
If $\phi_x \neq 0$, $k_z$ is not a constant of motion but $k_y$ and
$(\phi_z k_z + \phi_x k_x) / \sqrt{\phi_x^2 + \phi_z^2}$ are
constants of motion.

 The Fermi surface is
given by $k_x = \pm k_F$ in the zeroth order perturbation in $t_b /
t_a$ and $t_c / t_a$.
 When the condition
\begin{equation}
2 k_F =  m_z  u_x
      +m_y  u'_x
   + s (2\pi / a)
\label{ConditionU0}
\end{equation}
is satisfied with integers  $m_y$,
$m_z$ and $s$,
the degenerate states at $k_x = k_F$ and $k_x =
-k_F$ are mixed resulting
  the energy gap  at the
Fermi energy in the $|m_z|$th perturbation in  $(t_b /t_a)$ and
the $|m_y|$th perturbation in  $(t_c /t_a)$.  The magnitude of the
energy gap is of the order of $|\Gamma_0 ({\bf k})|$, where
  \begin{eqnarray}
 \Gamma_0 ({\bf k})
  &=&
   t_a \left( {t_b \over t_a } \right)^{ | m_z |}
       \left( {t_c \over t_a } \right)^{ | m_y |}
 e^{- i m_y \pi{\phi_x
 \phi_y \over \phi_0 \phi_z}}
\nonumber \\
 & & \quad   e^{-i m_z b k_y}
  \left( e^{  i  c k_z} e^{  i  c (k_z + {2\pi \over
c}{\phi_x\phi_y
 \over \phi0 \phi_z})} \cdots
 e^{  i  c (k_z + (|m_y| - 1)
   {2\pi \over c}{\phi_x\phi_y
 \over \phi_0 \phi_z})} \right)^{{\rm sign} (m_y)}
 \nonumber   \\
 &=&
   t_a \left( {t_b \over t_a } \right)^{ | m_z |}
       \left( {t_c \over t_a } \right)^{ | m_y |}
e^{- i m_y^2 \pi{\phi_x
\phi_y \over \phi_0 \phi_z}} e^{ i  ( - m_z b k_y + m_y c k_z) }  .
\label{G0}
\end{eqnarray}
Since the flux is assumed to be rational,
${\bf u}$ and ${\bf u}'$ are written as the integer multiple
of the vector
${\bf u}_0$ as
\begin{mathletters}
\begin{eqnarray}
 {\bf u} &=& P_z {\bf u}_0 ,   \\
 {\bf u}' &=& P_y {\bf u}_0 ,
\end{eqnarray}
\end{mathletters}
where
integers
$P_y$ and $P_z$ are defined by
\begin{equation}
 {P_y \over P_z} = {\phi_y \over \phi_z}.
\end{equation}
 The Harper equations are obtained by using the {\it
one dimensional} basis states
\begin{equation}
 \psi_j = \psi ( {\bf k} + j {\bf u}_0) ,
\end{equation}
with integer $j$.
 The number of the Harper equations obtained by this  basis set is
not always $Q$ but the integer multiple of $Q$, because $\psi_Q$ is
not necessarily the same as $\psi_0$.  Thus the basis  $\psi_j$ may
be overcomplete,  although the eigenenergy is obtained correctly by
the overcomplete basis. In order to get the complete basis  for the
rational flux case a  careful choice of the vector potential is
necessary\cite{HKM93}. The overcomplete basis
set corresponds to the larger unit cell in the real space or the
folded magnetic Brillouin zone in the momentum space. The Hall
conductivity is, however, obtained even if the integration
in  Eq.~(\ref{HallConductivity}) is done in the folded magnetic
Brillouin zone,  since  the degeneracy of the
bands due to the overcomplete bands compensates the folded region of
the integration.

 The magnetic  Brillouin
zone has the volume $(2\pi)^3/(Qabc)$ and $2 k_F  a Q / (2 \pi) = r$ is
the number of filled bands below the Fermi energy. Then we get from
Eq.~(\ref{ConditionU0}) that $m_y$ and $m_z$ should fulfill the
Diophantine equation,
 \begin{equation}
 r = sQ + {\phi_z \over \phi_0  } Q m_z + {\phi_y \over \phi_0  } Q
m_y.
\label{Diophantine}
\end{equation}

The wave function for the $r$th band gets the phase of
$\Gamma_0({\bf k})$ at $k_x \approx k_F$ \cite{Kohmoto89}, {\it
i.e.} the phase of the wave function cannot be defined globally in
the magnetic Brillouin zone. On the other hand the phase of the
wave function can be defined globally in the torus of $k_y$ and
$k_z$ for fixed $k_x$, which is the slice of three-torus by the plane
perpendicular to a fixed $k_x$.  Therefore, the second term in
Eq.~(\ref{HallConductivity}) is zero  by partial integration with
respect to $k_y$ or $k_z$ if $\beta = y$ or $\beta=z$.    Then the
conductivity tensor is written  as\cite{Kohmoto89,Yakovenko91}
\begin{mathletters} \label{Hallconductivityxyz}
 \begin{eqnarray}
\sigma_{x y }
&=&
  2 {1 \over c} {e^2 \over h } {1 \over 2 \pi i}
\int_{-\pi /b}^{\pi/b} dk_{y}
\left(
 <\Psi |  {\partial \over \partial k_{y}} | \Psi >
 \bigg|_{k_x=k_F^{-}}
  -
 <\Psi |  {\partial \over \partial k_{y}} | \Psi >
 \bigg|_{k_x=k_F^+}
\right)
,
  \label{Hallconductivityxy} \\
\sigma_{x z}
&=&
 2 {1 \over b} {e^2 \over h } {1 \over 2 \pi i}
\int_{-\pi/c}^{\pi/c} dk_{z}
\left(
 <\Psi |  {\partial \over \partial k_{z}} | \Psi >
 \bigg|_{k_x=k_F^{-}}
  -
 <\Psi |  {\partial \over \partial k_{z}} | \Psi >
 \bigg|_{k_x=k_F^+}
\right)
, \label{Hallconductivityxz} \\
\sigma_{yz} &=& 0 ,
\end{eqnarray} \end{mathletters}
where $k_F^{-}$ ( $k_F^+$ ) is the momentum smaller  (larger) than
the Fermi momentum.
Since the wave function changes only in the phase at the Fermi
momentum, the integration with respect to   $k_{y}$ or $k_z$ gives
the winding number of $\Gamma_0( {\bf k})$ in the complex plane
around zero when $k_y$ or $k_z$ is moved.  Using Eqs.~(\ref{G0}) and
(\ref{Hallconductivityxyz}), we obtain
\begin{mathletters}
\begin{eqnarray}
 \sigma_{xy} = {2\over c} {e^2 \over h} m_z, \\
 \sigma_{zx} =  {2\over b} {e^2 \over h} m_y.
\end{eqnarray}
\end{mathletters}
The conductivity per plane is quantized as $\tilde{\sigma}_{xy} =  2
(e^2/h)m_z$ and $\tilde{\sigma}_{zx} = 2
(e^2/h)m_y$.

  Unfortunately, the energy
gap  will be very small if
 $|m_y| + |m_z| \gg 1$, which is the case for $|\phi_y|$, $|\phi_z|
\ll \phi_0$.

For the quantum Hall effect, it is not necessary that all  of
$\phi_x/\phi_0$, $\phi_y/\phi_0$ and $\phi_z/\phi_0$ are integers.
The only required condition is Eq.~(\ref{ConditionU0}) .

\section{3D Quantum Hall effect in the presence of FISDW}
\label{fisdw}
In this section we take account of the interaction
\begin{equation}
{\cal H}'
=  U \sum_{i} c_{i,\uparrow}^{\dagger} c_{i,\uparrow}
                     c_{i,\downarrow}^{\dagger} c_{i,\downarrow} .
\end{equation}
 This interaction is written as
\begin{equation}
  {\cal H}'
  =  -  U  \left( {a b c \over (2 \pi)^3 } \right)^3
   \int d{\bf k}
   \int d{\bf k}'
   \int d{\bf K} \
 c_{\uparrow}^{\dagger} ({\bf k})
   c_{\downarrow} ({\bf k} + {\bf K})
   c_{\downarrow}^{\dagger} ( {\bf k}' + {\bf K})
   c_{\uparrow} ( {\bf k}')  .
  \label{eqU}
\end{equation}
The order
parameter of the spin density is defined in the mean field
approximation as
\begin{equation}
  \Delta ({\bf K})
   \equiv -U  {a b c \over (2 \pi )^3 }
    \int d{\bf k}'
     < c_{\downarrow}^{\dagger} ( {\bf k}' + {\bf K})
       c_{\uparrow}   ( {\bf k}')  >.
\label{OrderParameter}
\end{equation}
The interaction term mixes the states
$\psi_{\uparrow} ({\bf k})$ and $\psi_{\downarrow} ({\bf k} + {\bf
K})$. As a result
  an energy
gap at the Fermi surface is opened by the order parameter when
 $K_x=\pm 2 k_F$.  The $y$ and
$z$ components of the wave vector should be determined to give the
lowest energy, which can be calculated numerically with further
approximation such as the linearization with respect to $k_x$
\cite{GorkovLebed84,Montambaux85,Virosztek86}.
 Since we focus on the
quantization of the Hall conductivity,  we do not have to
calculate the  values of $K_y$ and $K_z$ explicitly.
The Hall conductivity is zero in this case.

There are other possibilities for opening
gaps in the presence of the magnetic field.  The order parameter
may have many components with respect to the wave vectors as in the
two-dimensional case \cite{Machida93,Hori93,Machida94}. In
three-dimensional case the order parameter has the form
 \begin{equation}
 \Delta ({\bf K}) = \sum_{n_y, n_z} D_{n_y,n_z} \delta ({\bf K} -
{\bf K}_{n_y,n_z}),
 \end{equation}
with the
$x$ component of the wave vector ${\bf K}_{n_y,n_z}$ given by
\begin{eqnarray}
 ({\bf K}_{n_y,n_z})_x &=& 2 k_F + n_y u'_x + n_z u_x
\nonumber \\
&=& 2 k_F + n_y {2 \pi \over a} {\phi_y \over \phi_0}
+ n_z {2 \pi \over a} {\phi_z \over \phi_0} ,
\label{WaveVector}
\end{eqnarray}
 where $n_y$ and $n_z$ are integers. The vectors
${\bf u}$ and ${\bf u}'$ are given in Eqs.~(\ref{VectorU}) and
(\ref{VectorUp}).
 By this wave vector the
state at $k_x=-k_F$ is mixed with that at  $k_x=k_F + n_y u_x'
+ n_z u_x$.
The state at $k_x=k_F + n_y u_x' + n_z
u_x$  is mixed with the state at $k_x=k_F$ by $| n_y |$th
perturbation in $t_b \exp (ibk_y)/t_a$ and $| n_z |$th perturbation
in $t_c \exp(ick_z)/t_a$. As a result  two states at the Fermi energy
at $k_x = \pm k_F$ are mixed by a combined effect of SDW and the
magnetic field. (See Fig. 1.) The gap function is given by
\begin{equation}
 \Gamma( {\bf k} ) = \sum_{n_x,n_y} \Gamma_{n_y,n_z}
\exp \left[ i  \left( - n_z b k_y +  n_y c k_z - \pi n_y^2 {\phi_x
\phi_y \over \phi_z \phi_0} \right) \right],
\label{GapFunction}
\end{equation}
where $\Gamma_{n_y,n_z} $ is a constant of the order of
$D_{n_y,n_z}  ( t_b / t_a ) ^{|n_z|} (t_c / t_a ) ^{|n_y|}$.
The  wave function near the Fermi momentum changes the phase due
to this gap function. The Hall conductivity is calculated by
Eq.~(\ref{HallConductivity}) as in the non-interacting case.

If only one component of the order parameter is dominant,
{\it i.e.},
\begin{equation}
 \Delta({\bf K}) \approx D_{n_{y0},n_{z0}} \delta ({\bf K} -
{\bf K}_{n_{y0},n_{z0}}) ,
\end{equation}
the $k$-dependent phase change is $\exp [i (-n_{z0} b k_y  + n_{y0}
c k_z)]$ and the Hall conductivity is obtained by
Eqs.~(\ref{Hallconductivityxyz}) as \begin{mathletters}
\label{sigma2}
\begin{eqnarray}
 \sigma_{xy} &=& {2\over c} {e^2 \over h} n_{z0} , \label{sigma2a}\\
 \sigma_{zx} &=& {2\over b} {e^2 \over h} n_{y0},
\end{eqnarray}
and
\begin{equation}
\sigma_{yz} = 0.  \label{sigma2c}
\end{equation}
\end{mathletters}
Comparing Eqs.~(\ref{sigma2a}) - (\ref{sigma2c}) with
Eqs.~(\ref{sigma1}) and (\ref{defG2}) we find that
\begin{mathletters}
\begin{eqnarray}
l_a &=& 0  ,   \\
l_b &=& n_y ,  \\
l_c &=& n_z  .
\end{eqnarray}
\end{mathletters}
If the order parameter consists of  many components  with respect
to the wave vectors $K_{n_y,n_z}$, the Hall conductivity
$\sigma_{xy}$ ($\sigma_{xz}$) is obtained as a winding number of the
gap function $\Gamma (\bf k)$ as $k_y$ ($k_z$) is moved  from $-\pi
/ b$ to $\pi / b$  (from $-\pi / c$ to $\pi / c$).
Note that $\phi_x$ appears as the relative phase between the
components of the order parameter, so the Hall conductivity may
depend on $\phi_x$ implicitly.

If the magnetic field is changed
in magnitude or direction by a small amount, the wave vectors and
the order parameter will  change in order to keep the Fermi energy in
the same energy gap. As a result the plateaux of the Hall
conductivity is realized without the effect of a localization. When
the energy gap is closed and opens again at the Fermi energy as the
magnetic field is changed, the Hall conductivity changes. The sign
change of the Hall conductivity may happen as the direction or the
magnitude of the magnetic field is changed.

The condition for the rational flux is not necessary to get the
Quantum Hall effect.  The wave vector of the order  parameter is
adjusted to satisfy Eq.~(\ref{WaveVector}) for any direction and
amplitude of the magnetic field.  If $\phi_y / \phi_z$ is
irrational,  $n_y$ and $n_z$ are determined uniquely for a given
wave vector $({\bf K}_{n_y, n_z})_x$.  On the other hand $n_y$ and
$n_z$ are not uniquely determined if $\phi_y / \phi_z$ is rational.

\section{triclinic lattice}
\label{triclinic}
In this section we consider the triclinic lattice with ${\bf a}$,
${\bf b}$ and ${\bf c}$  not being orthogonal each other.
The fundamental
reciprocal vectors,
\begin{mathletters}
\begin{eqnarray}
{\bf G}_a &=& 2\pi {{\bf b} \times {\bf c} \over v} \\
{\bf G}_b &=& 2\pi {{\bf c} \times {\bf a} \over v} \\
{\bf G}_c &=& 2\pi {{\bf a} \times {\bf b} \over v},
\end{eqnarray}
\end{mathletters}
are not orthogonal,
where
\begin{equation}
 v = |{\bf a} \cdot ({\bf b} \times {\bf c})  |.
\end{equation}
We take
${\bf a}  \parallel {\bf x}$.
 Hopping matrix elements $t_a$, $t_b$
and $t_c$ ($|t_a| \gg |t_b|$,~$|t_c|$) are assumed between the
nearest sites along $a$, $b$ and $c$ axes.  The vector potential is
taken as Eq.~(\ref{VectorPotential})  by using the orthogonal $x$,
$y$ and $z$ axes.

The non-interacting Hamiltonian is
\begin{eqnarray}
{\cal H}_0 = &-& t_a \sum_{{\bf k}, \sigma} 2\cos ({\bf a} \cdot
{\bf k}) c^{\dagger}({\bf k}) c({\bf k})
\\ \nonumber
 &-& t_b e^{i \theta_b} \sum_{{\bf k}, \sigma} e^{-i {\bf b}
\cdot {\bf k}}
 c^{\dagger}_{\sigma} ({\bf k} - {\bf w}) c_{\sigma} ({\bf k} )
+ h.c.
\\ \nonumber
&-& t_c e^{i \theta_c} \sum_{{\bf k}, \sigma} e^{-i {\bf c}
\cdot {\bf k}}
 c^{\dagger}_{\sigma} ({\bf k} + {\bf w}') c_{\sigma} ({\bf k} ) +
h.c. ,
 \end{eqnarray}
where
\begin{mathletters}
\begin{eqnarray}
\theta_b &=& {\pi \over \phi_0}
(b_y - {B_y \over B_z} b_z)(B_z b_x - B_x b_z) \\
\theta_c &=& {\pi \over \phi_0}
(c_y - {B_y \over B_z} c_z)(B_z c_x - B_x c_z)
 \end{eqnarray}
and
\end{mathletters}
\begin{mathletters}
\begin{eqnarray}
{\bf w} = {2\pi \over \phi_0} (b_y - {B_y \over B_z} b_z) (B_z
\hat{\bf x} - B_x \hat{\bf z}), \\
{\bf w}' = - {2\pi \over \phi_0} (c_y - {B_y \over B_z} c_z) (B_z
\hat{\bf x} - B_x \hat{\bf z}).
 \end{eqnarray}
\end{mathletters}
Note that ${\bf w} \parallel {\bf w}'$.

The Fermi surface in the
zeroth order in $t_b/t_a$ and $t_c/t_a$ is the plains given by
${\bf a}\cdot{\bf k} = \pm a k_F$.  The
order parameter is defined by Eq.~(\ref{OrderParameter}) with the
factor $v/(2 \pi)^3$ instead of $abc/(2 \pi)^3$. As in the previous
section, the energy gap opens at the Fermi momentum by the effects of
the order parameter and the perturbation  in $t_b/t_a$ and $t_c/t_a$, if
the order parameter has the form \begin{equation}
\Delta({\bf K}) = \sum_{n_b,n_c} D_{n_b,n_c}\delta({\bf K} - {\bf
K}_{n_b,n_c}), \label{OrderParameter3D}
\end{equation}
where $n_b$ and $n_c$ are integers and the
 wave vector ${\bf K}_{n_b,n_c}$ satisfies the condition
\begin{eqnarray}
 {\bf K}_{n_b,n_c} \cdot {\bf a} &=& 2 a k_F + n_b {\bf w}'
\cdot {\bf a} +  n_c {\bf w}  \cdot {\bf a}
\nonumber \\
&=& 2 a k_F + n_b 2\pi {a c_y B_z - a c_z B_y \over \phi_0}
  - n_c 2\pi {a b_y B_z - a b_z B_y \over \phi_0}
\nonumber \\
&=& 2 a k_F + n_b 2\pi {\phi_b \over \phi_0}
  + n_c 2\pi {\phi_c \over \phi_0}.
\end{eqnarray}
In the above $\phi_b$ and $\phi_c$ are the fluxes through a unit
area in $a$-$c$ and $a$-$b$ planes, respectively,
\begin{mathletters}
\begin{eqnarray}
\phi_c &=&   {\bf a} \cdot ({\bf b} \times {\bf B}   )\\
\phi_b &=& - {\bf a} \cdot ({\bf c} \times {\bf B}   )
\end{eqnarray}
\end{mathletters}
The  phase of the wave function is not defined globally as in the
previous sections. In the perturbation in $t_b/t_a$ and $t_c/t_a$,
the mismatch of the locally defined phase for the wave function of
the highest occupied band is attributed to the phase of the gap at
the Fermi momentum. The wave function changes
the phase  as the momentum is moved at the Fermi momentum as in the
previous section.
We consider the case that  one component is
dominant in Eq.~(\ref{OrderParameter3D}). Then the
${\bf k}$-dependent part of the phase change is
  \begin{equation}
\exp \left[ i \left( -n_c {\bf b} \cdot{\bf k} + n_b {\bf c} \cdot
{\bf k} \right) \right].
 \end{equation}

The conductivity tensor is calculated by
\begin{mathletters}
\begin{eqnarray}
\sigma_{x y }
&=&
  2  {e^2 \over h } {1 \over 2 \pi^2 i}
\int dk_{y} dk_{z}
\left(
 <\Psi |  {\partial \over \partial k_{y}} | \Psi >
 \bigg|_{k_x=k_F^{-}}
  -
 <\Psi |  {\partial \over \partial k_{y}} | \Psi >
 \bigg|_{k_x=k_F^+}
\right)
,
  \label{Hallconductivityxy_ani} \\
\sigma_{z x}
&=&
 - 2  {e^2 \over h } {1 \over 2 \pi^2 i}
\int dk_{y} dk_{z}
\left(
 <\Psi |  {\partial \over \partial k_{z}} | \Psi >
 \bigg|_{k_x=k_F^{-}}
  -
 <\Psi |  {\partial \over \partial k_{z}} | \Psi >
 \bigg|_{k_x=k_F^+}
\right)
, \label{Hallconductivityxz_ani} \\
\sigma_{yz} &=& 0 .
\end{eqnarray}
\end{mathletters}
  Because of the non-orthogonality of ${\bf a}$, ${\bf b}$ and
${\bf c}$, the slice of the magnetic Brillouin zone by  the plane
perpendicular to  $k_y$ or $k_z$ is not the torus. As a result,
$\sigma_{x y}$ and $\sigma_{z x}$ are not  quantized to be integer
but they are quantized as \begin{eqnarray}
\sigma_{x y } &=& 2 {e^2 \over h} (n_c b_y - n_b c_y) { a \over
v} \nonumber \\
&=& 2 {e^2 \over h} { n_c b_y - n_b c_y \over |b_y c_z - b_z c_y|}
\\
\sigma_{z x} &=& 2 {e^2 \over h} (-n_c b_z + n_b c_z) { a \over
v}
\nonumber \\
&=& 2 {e^2 \over h} {-n_c b_z + n_b c_z \over |b_y c_z - b_z c_y|}.
\end{eqnarray}
In the above we have used the facts that ${\bf a} \parallel {\bf x}$
and the area of the Brillouin zone in $k_y$-$k_z$ plane is
$(2 \pi)^2 a / v$.  Comparing these result with the general
expression of the Hall conductivity in three-dimensional periodic
system (Eq.~(\ref{defG2})), we get \begin{mathletters}
\begin{eqnarray}
l_a &=& 0 \\
l_b &=& n_b \\
l_c &=& n_c.
\end{eqnarray}
\end{mathletters}

\section{conclusion}
When the magnetic field is tilted to any direction,
there is a possibility for the new wave vector for FISDW.  We have
shown that the Hall conductivity $\sigma_{xy}$, $\sigma_{zx}$ and
$\sigma_{yz}$ are  quantized as a function of the wave number of
FISDW. When the lattice is orthogonal, both $\sigma_{xy}$ and
$\sigma_{xz}$ per plane are quantized as integers and $\sigma_{yz}$
stays to be zero as long as $t_b/t_a$ and $t_c / t_a$ are treated in
perturbation. Since the quantization is realized with respect to the
fundamental reciprocal lattice, $\sigma_{xy}$ and $\sigma_{zx}$ is
not quantize as integers if the ${\bf a}$, ${\bf b}$ and ${\bf c}$
axes are not orthogonal each other. Even in that case the quantum
values of $\sigma_{xy}$ and $\sigma_{zx}$ are given as a function of
the wave vector of FISDW, which is characterized by  integers.

\bigskip
\acknowledgments

The author would like to thank M. Kohmoto for helpful discussions.

%
 \begin{figure}
 \caption{
Schematic diagram of opening the gap at the Fermi momentum. The
right-arrow labeled by $D_{0,0}$ indicates the order parameter of
SDW with $K_x = 2 k_F$ by which two states at $k_x = \pm k_F$ are
coupled. The right-arrow labeled by $D_{1,1}$ indicates the SDW with
wave vector $K_x = 2 k_F + u_x + u_x'$. The left-arrows indicate the
coupling of the states  caused by the magnetic field in the
perturbation in $t_b/t_a$ and $t_c / t_a$.
 }
 \label{fig1}
\end{figure}

\end {document}